\documentclass[aps,prl,twocolumn,showpacs]{revtex4}
\usepackage{amssymb}
\usepackage{amsmath, amsthm}

\begin{document}


\title{A 
Lema{\^{\i}}tre--Tolman--Bondi cosmological wormhole}



\author{I. Bochicchio}
\email[]{ibochicchio@unisa.it} 
\affiliation{Dipartimento di Matematica e Informatica, 
Universit\'a degli Studi di Salerno, Via Ponte Don Melillo, 84084 
Fisciano (SA), Italy}

\author{Valerio Faraoni}
\email[]{vfaraoni@ubishops.ca}
\affiliation{Physics Department, Bishop's University\\
Sherbrooke, Qu\'ebec, Canada J1M~1Z7
}


\begin{abstract}
We present a new analytical solution of the Einstein field
equations describing a wormhole shell of zero thickness joining
two Lema{\^{\i}}tre--Tolman--Bondi universes, with no radial
accretion. The material on the  shell satisfies the energy
conditions and, at late times, the  shell becomes comoving with
the dust-dominated cosmic substratum.
\end{abstract}

\pacs{04.20.-q, 04.20.Jb, 98.80.-k }
\keywords{Lema{\^{\i}}tre--Tolman--Bondi solutions; cosmological
wormholes}

\maketitle

\section{Introduction}

Static and asymptotically flat wormhole solutions of the Einstein
equations have been known for a long time \cite{staticwormholes}. 
The study of wormholes developed with the seminal paper by Morris
and Thorne \cite{MorrisThorne}, after which many solutions were
discovered (see \cite{Visserbook} for an extensive discussion).
The possibility that inflation in the early universe may enlarge a
Planck size wormhole to a macroscopic size object was contemplated
in Ref.~\cite{Roman93}. Dynamical wormholes were discovered and
studied in Refs.~\cite{HochbergVisser98, Hayward99} and wormholes
in cosmological settings were contemplated in various works
(\cite{Kim96, othercosmowormholes} and references therein), with
particular attention being paid to wormholes with cosmological
constant $\Lambda$, which are asymptotically de Sitter or anti--de
Sitter according to the sign of  $\Lambda$  
\cite{LemosLoboOliveira03}.

After the 1998 discovery of the present acceleration of the
universe \cite{SN} and the introduction of dark energy in
cosmological theories to account for this cosmic acceleration, 
there were claims that phantom energy, an extremely exotic form of
dark energy with $P < -\rho$ (where $P$ and $\rho$ are the
pressure and the energy density, respectively) could cause the
universe to end with a Big Rip singularity at a finite time in the
future \cite{BigRip}.  There was then a claim in the literature
\cite{GonzalezDiaz04} that, if a wormhole accretes phantom energy,
it grows to enormous size faster than the background universe,
swallowing the entire cosmos  which would then tunnel through the
wormhole throat and re-appear in a different portion of the
multiverse before reaching the Big Rip singularity. This claim was
based on qualitative arguments  and was later disproved by two
classes of exact solutions of the Einstein equations representing
wormholes embedded in a cosmological background dominated by
phantom energy \cite{FaraoniIsrael05}. These wormholes accrete
phantom energy but, even if their expansion rate differs from that
of the cosmic substratum initially, they become comoving with it
as the scale factor of the
Friedmann--Lema{\^{\i}}tre--Robertson--Walker (FLRW) universe in
which they are embedded grows.

The first class of solutions consists of a zero-thickness shell
which carries exotic energy and does not perturb the two copies of
the FLRW universe which it joins.  The second, more realistic, 
class
is described by a generalized McVittie metric \cite{McVittie} with
an imperfect fluid and a radial energy flow, with the  mass of
the wormhole shell distorting the surrounding FLRW  
metric \cite{FaraoniIsrael05}. Another, less general, solution of 
the
Einstein equations describing a cosmological wormhole comoving
with the background was presented in Ref.~\cite{SushkovKim04}.

Cosmological wormholes are truly dynamical and interest in this
kind of solution has developed in parallel with the increasing
attention paid to  cosmological black 
holes \cite{cosmologicalblackholes}.  Additionally, gravitational 
lensing
by wormholes was studied in \cite{wormholelensing} and numerical
solutions interpreted as wormholes in accelerating FLRW universes
were presented in Refs.~\cite{MaedaHaradaCarr08}.  Recently,
Maeda, Harada, and Carr have given precise definitions for general
cosmological wormholes and have found two new exact solutions of
this kind \cite{MaedaHaradaCarr09}. An important result of this
work, which echoes a previous result of \cite{Kim96}, is that the
null energy condition needs not be violated in this dynamical
situation, although it must be violated for static wormholes to
exist \cite{MaedaHaradaCarr09}. It seems that the study of
cosmological wormholes is developing into a promising new area of
research.

In this paper we propose a new analytical solution of the Einstein
field equations describing a cosmological wormhole shell joining
two Lema\^{\i}tre--Tolman--Bondi (LTB) universes. We are led to this
solution by the following considerations: the second class of
solutions in Ref.~\cite{FaraoniIsrael05} is inspired by the
McVittie  metric, which describes a central inhomogeneity in a
FLRW  background. However, the McVittie metric  needed to be
generalized by removing the McVittie ``no accretion'' condition
$G_{01}\,=\,0$ (in spherical coordinates) which forbids radial
energy flow. The goal of Ref.~\cite{FaraoniIsrael05} was to
describe the effect of the accretion of phantom energy onto the
wormhole. Here, we begin by noting  that inhomogeneities embedded
in a FLRW background are usually described by using an LTB 
metric \cite{Lemaitre, Tolman, Bondi}, not a McVittie one. The 
classical
LTB metric describes a spherically symmetric inhomogeneity in a
dust--dominated FLRW background. The Bondi condition $G_{01}= 0 $
parallels the McVittie no-accretion condition and forbids the
(radial) flow of cosmic dust onto the inhomogeneity. It would be
interesting to obtain a solution describing a wormhole shell
joining two identical LTB universes and perturbing its
surroundings in the way described by the LTB metric. This is what
we do here. We obtain a wormhole shell composed of exotic matter
which expands more slowly than the cosmic substratum (which
becomes a spatially flat FLRW universe at late times), but
eventually becomes comoving with it.

The next section details how to construct the wormhole shell and
satisfy the Israel--Darmois--Lichnerowicz junction conditions
\cite{junction} on this shell. The Einstein equations on the shell
provide expressions for the energy density and pressure of the
material on the shell. Sec.~III uses the covariant conservation
equation to relate the  rate of change of the mass of shell
material, the shell area, and the flux of cosmic fluid onto the
shell due to the relative velocity between the shell and the
cosmic substratum. The metric signature is $-+++$, we use units in
which the speed of light and Newton's constant are unity, and we 
follow the notations of
Ref.~\cite{Wald}. Greek indices run from~0 to~3 and 
Latin indices assume the values $0, 1$, and $2$ corresponding to 
the coordinates $\left( t, \theta, \varphi \right)$ of the 
spherical hypersurface $\Sigma$ defined below.

\section{The LTB wormhole solution}

The spherically symmetric LTB line element for the critically 
open universe in polar coordinates $
\left( t, r, \vartheta, \varphi \right)$ is
\begin{equation}\label{metric}
ds^2\,=\,-dt^{2}+\left[ R^{\prime }(t,r)\right]
^{2}dr^{2}+R^{2}\left( t,r \right) \, d\Omega ^{2}\ ,
\end{equation}
where
\begin{equation}\label{raggio}
R\left( t,r \right) = \left( r^{3/2}+\frac{3}{2} \,
\sqrt{m_{e}(r)}\,\,t\right) ^{2/3}
\end{equation}
is an areal radius, $r$ is a comoving radius,
\begin{equation}\label{massa}
m_{e}(r)=4\pi \int\limits_{0}^{r}dx \, x^{2}\rho _{0}(x) \,,
\end{equation}
$\rho _{0}(r)$ is the energy density on an initial hypersurface, a
prime denotes differentiation with respect to $r$, and $d\Omega
^{2} \equiv d\vartheta ^{2}+\sin ^{2}\vartheta \,d\varphi ^{2}$. 
The line
element \eqref{metric} describes a spherical inhomogeneity in a
dust--dominated universe (\cite{Lemaitre, Tolman, Bondi}; for a
recent review see \cite{IvanaFrancavigliaLaserra09}).

Consider now a wormhole shell $\Sigma$ at $r\,=\,r_{\Sigma}(t)$
which joins two identical copies of an LTB spacetime (this shell
describes a wormhole created with the universe and not formed as
the result of a dynamical process after the Big Bang). The 
wormhole shell is dynamical and moves, possibly also relative to 
the cosmic substratum, and its motion is described by the form of 
the function $r_{\Sigma}(t)$. It is
convenient to write the equation of this shell as \footnote{Since
$r$ is a comoving radius, if $r_{\Sigma }$ were constant, the
shell would be comoving. But $r_{\Sigma}$ depends on time,
hence a relative motion between the shell and the cosmic
background is allowed.  However, we will find below that the shell
becomes comoving at late times. Moreover, one could define 
$N_{\mu}$ as $\nabla_{\mu} \left( r-r_{\Sigma}(t) 
\right)$ instead  but this, of course, leads to the same unit 
normal $n_{\mu}$.}
\begin{equation}\label{shellequation}
f\left( t, r \right) \equiv R-R_{\Sigma }\left( t,\,r_{\Sigma
}(t)\, \right)=0   \,.
\end{equation}
To find the unit normal to $\Sigma$ we first compute
\begin{equation}\label{normalebassa}
N_{\mu } \equiv \nabla _{\mu}f =\nabla _{\mu } \left( R-R_{\Sigma
} \right)=\left( R_{t}-\dot{{R}}_{\Sigma },R^{\prime }\,,0,0
\right)\,,
\end{equation}
and
\begin{equation}\label{normalealta}
N^{\mu }=\left( \dot{R}_{\Sigma }-R_{t}\,,\frac{1}{R^{\prime
}}\,,0,0 \right) \,,
\end{equation}
and then normalize according to $n_{\mu}\,=\,\alpha\,N_{\mu}$. 
Here $R_t \equiv \partial R/\partial t$ and an overdot denotes a 
total derivative with respect to $t$, {\em i.e.}, $\dot{R}\equiv 
dR/dt$.

The normalization $n_{\mu}n^{\mu}\,=\,1$ yields
\begin{equation}\label{parametro}
\alpha =\frac{1}{\sqrt{1-\left(
R_{t}-\dot{R}_{\Sigma}\right)^{2}}} \ .
\end{equation}
It is convenient to introduce the radial velocity of the wormhole
shell relative to the cosmic substratum
\begin{equation}\label{relativelocity}
v\equiv \,\dot{R}_{\Sigma }-R_{t\mid \Sigma } \,,
\end{equation}
where $R_{t\mid \Sigma}\equiv R_t \left( t, r_{\Sigma}(t)
\right)$. Then
\begin{equation}
\alpha\,=\,\frac{1}{\sqrt{1-v^2}}\,=\,\gamma(v)
\end{equation}
is an (instantaneous \footnote{It is not assumed that $v$ is a
constant.}) Lorentz factor for the relative motion
shell--background. The unit normal is then
\begin{eqnarray}
n_{\mu } & = & \left( -\gamma \,v,\gamma \,R^{\prime }\,,0,0
\right) \,, \\
&&\nonumber\\
n^{\mu } & = & \left( \gamma \,v,\frac{\gamma }{R^{\prime
}}\,\,,0,0 \right) \,.
\end{eqnarray}
The restriction of the metric to $\Sigma$ is given by
\begin{equation}
ds_{\mid \Sigma }^{2}=-dt^{2}+R^{\prime}_{\Sigma}dr_{\mid \Sigma
}^{2}+R_{\Sigma }^{2}d\Omega ^{2}
\end{equation}
or, using the fact that $\dot{R}_{\Sigma }=R_{t\mid \Sigma
}+R_{\Sigma }^{\prime }\,\dot{r}_{\Sigma }$ on the shell,
\begin{equation}
ds_{\mid \Sigma }^{2}= -(1-v^{2})dt^{2}+R_{\Sigma }^{2}d\Omega
^{2} \,,
\end{equation}
which expresses the fact that the proper time $\tau$ of the shell
is given by
\begin{equation}
d\tau\,=\,\sqrt{1-v^2} \, \,dt\,,
\end{equation}
\emph{i.e.}, it is Lorentz-dilated with respect to the comoving
time $t$ of the background.

Using the triad
\begin{equation}
\left\{ e_{(t)}^{\alpha } \,, e_{(\vartheta )}^{\alpha} \,,
e_{(\varphi )}^{\alpha} \,\, \right\} =\left\{ \sqrt{1-v^{2}}
\,\delta^{\alpha t}, \delta^{\alpha \vartheta },\delta^{\alpha
\varphi } \right\} \,,
\end{equation}
the extrinsic curvature of the shell is given by $(a, b, c=t,
\vartheta\,\,, \text{or}\,\,\varphi)$
\begin{equation} \label{curvature}
K_{\alpha \beta }=e_{\alpha }^{\left( a\right) }\,e_{\beta
}^{\left( b\right) }\,\nabla _{a}n_{b}=e_{\alpha }^{\left(
a\right) }\,e_{\beta }^{\left( b\right) }\left( \partial
_{a}n_{b}-\Gamma _{ab}^{c}n_{c}\right)\,,
\end{equation}
where $\Gamma _{ab}^{c}$ are the Christoffel symbols of the
$3$--dimensional metric $g_{ab\mid \Sigma }$.
Eq.~\eqref{curvature} yields
\begin{equation}
\begin{array}{lll}
K_{tt}=\frac{1}{\gamma ^{2}}\left( \partial _{t}n_{t}-\Gamma
_{tt}^{t}n_{t}\right) =-\frac{v_{t}}{\gamma }-2\gamma
v_{t}v^{2}\,,
\\[1em]
K_{\vartheta \vartheta }=-\Gamma _{\vartheta \vartheta
}^{t}n_{t}=\gamma ^{3}v\,R_{\Sigma }R_{t\mid \Sigma }\,,
\\[1em]
K_{\varphi \varphi }=K_{\vartheta \vartheta }\,\sin ^{2}\vartheta
\,.
\end{array}
\end{equation}
The mixed components of the extrinsic curvature are
\begin{eqnarray}
{K^{t}}_{t} & = & \gamma \, v_{t}\left( 2\gamma
^{2}v^{2}+1\right) \,,\\
&&\nonumber\\
{K^{\vartheta }}_{\vartheta } & = & \gamma ^{3}v\,\frac{R_{t\mid
\Sigma  }}{R_{\Sigma }} = {K^{\varphi }}_{\varphi }\,,
\end{eqnarray}
while the trace is
\begin{equation}
K={K^{t}}_{t}+ { K^{\vartheta }}_{\vartheta } + { K^{\varphi
}}_{\varphi }= 2 \gamma ^{3}v\,\left( \frac{R_{t\mid \Sigma
}}{R_{\Sigma }}+v_{t}v\right) +\gamma v_{t} \,.
\end{equation}
Since there are two identical LTB universes joining at the shell
with unit normal $n^{\mu}$ pointing outward, the jumps of these
quantities on $\Sigma$ are
\begin{equation}
\left[ K_{\,\,\,\,\,\,b}^{a}\right] =2K_{\,\,\,\,\,\,b}^{a} \,,
\quad \quad \left[ K\right] =2K \,.
\end{equation}
The Einstein equations at the shell $\Sigma$ are
\cite{BarrabesIsrael91}
\begin{equation}\label{EES}
\left[ K_{\,\,\,b}^{a}-\delta _{\,\,\,b}^{a}\,K\right] =-8\pi
\,S_{\,\,\,b}^{a} \,,
\end{equation}
where $S_{ab}$ is the energy--momentum tensor of the material on
the shell. We assume that this matter is a perfect fluid,
described by
\begin{equation}
S_{ab}=\left( \sigma +P_{\Sigma }\right) u_{a}^{\left( \Sigma
\right) } \, u_{b}^{\left( \Sigma \right) }+P_{\Sigma
}\,\,g_{ab\mid \Sigma }\,,
\end{equation}
where $\sigma$ and $P_{\Sigma}$ are the $2$--dimensional surface
density and pressure, respectively, and $u_{\left( \Sigma \right)
}^{\mu }$ is the $4$--velocity of the shell given by
\begin{equation}
u_{\left( \Sigma \right)}^{\alpha }=\frac{d\,x_{\left( \Sigma
\right) }^{\alpha }}{d\tau }=\frac{\partial x_{\left( \Sigma
\right) }^{\alpha }}{\partial x^{\mu }}\frac{dx^{\mu }\,}{d\tau
}=\gamma\,\,\frac{dx^{\mu }\,}{dt} \, \frac{\partial x_{\left(
\Sigma \right) }^{\alpha }}{\partial x^{\mu }}\,.
\end{equation}
The coordinates on $\Sigma$ are $x_{\Sigma }^{\mu }=\left(
t,\,r_{\Sigma }(t),\,\vartheta ,\,\varphi \right)$, which yield
\begin{equation}
u_{\left( \Sigma \right) }^{\alpha }=\left( \gamma \,,\,\gamma
\,\frac{v}{R_{\Sigma }^{\prime }\,\,} \,\,,\,0\,,0\right) \,,
\;\;\,\,\, u_{\alpha }^{\left( \Sigma \right) }=\left( -\gamma
\,,\,\gamma \,v\,R_{\Sigma }^{\prime }\,,\,0\,,0\right) \,.
\end{equation}
As such, it is easy to see that
\begin{equation}
u_{\left( \Sigma \right) }^{\mu }\, u_{\mu }^{\left( \Sigma
\right) } =1 \,, \quad\quad u_{\left( \Sigma \right) }^{\mu
}\,n_{\mu }=0\,.
\end{equation}
The $\left(t, t \right)$ component of the Einstein
equations~\eqref{EES}  at  the shell is
\begin{equation}
\sigma =-\gamma ^{3} \,\frac{v}{2\pi }\,\frac{R_{t\mid \Sigma
}}{R_{\Sigma }}\,,
\end{equation}
while the $ \left(\vartheta, \vartheta \right)$ or the $
\left(\varphi, \varphi \right) $ component yields
\begin{eqnarray}
P_{\Sigma } & = & \frac{\gamma }{4\pi }\left( v_{t}+2\gamma^{2}
v^{2}v_{t}+\gamma ^{2}v\frac{R_{t\mid \Sigma }}{R_{\Sigma
}}\right) \nonumber\\
&&\nonumber\\
& = &  -\frac{\sigma }{2}+\gamma 
\frac{v_{t}}{4\pi } 
\frac{1+v^{2}}{1-v^{2}} \,.
\end{eqnarray}
Using eq.~\eqref{raggio}, one obtains
\begin{equation}
\frac{R_{t}}{R}=  \frac{\sqrt{m_{e}\left(
r\right)} }{ r^{3/2} + 3\sqrt{m_{e}\left( r\right)
}\,\,t /2 } \,;
\end{equation}
this quantity is asymptotic to $\frac{2}{3 t}$, the Hubble
parameter of the dust--dominated cosmological background, as $t
\rightarrow +\infty$. It is also
\begin{equation}
\sigma =-\gamma ^{3} \frac{v}{2\pi }\frac{R_{t\mid \Sigma
}}{R_{\Sigma }}=  -\gamma ^{3} \frac{v}{2\pi }\, 
\frac{\sqrt{m_{e}\left( r_{\Sigma}\right)} }{r_{\Sigma
}^{3/2}+3\sqrt{m_{e}\left( r_{\Sigma }\right) }\,\, t/2}
\end{equation}
and $\sigma >0$ is equivalent to $v<0$. A wormhole shell with
positive surface energy density must necessarily expand slower
than the cosmic substratum, a fact that is interpreted as the
influence of the inhomogeneity slowing down the expansion locally.
Since the expansion rate of the background $ \frac{2}{3t}$ tends
to zero at late times, the shell expansion rate must also tend to
zero and the shell becomes comoving. In fact, since $v<0$, it is
$\dot{R}_{\Sigma }=R_{t\mid \Sigma }+R_{\Sigma }^{\prime } \,
\dot{r}_{\Sigma }<R_{t\mid \Sigma }$ and, since $R_{\Sigma
}^{\prime }>0$, it is  $\dot{r}_{\Sigma }<0$. This inequality is
consistent, of course, with the relation
$$\dot{r}_{\Sigma }=\frac{v}{R_{\Sigma }^{\prime }}$$
which is easy to derive.

Now, $r_{\Sigma }>0$ decreases monotonically implying
that \footnote{We assume that the function $r_{\Sigma }(t)$ is
continuous.}, as $t \rightarrow +\infty$, either $r_{\Sigma }$
tends to a horizontal asymptote $r_{\infty}>0$, or
$\lim_{t \to +\infty}r_{\Sigma }(t)=0^{+}$. If $r_{\Sigma 
}\rightarrow 0^{+}$, then
\begin{equation}R_{\Sigma }=\left( r_{\Sigma
}^{3/2}+\frac{3}{2}\sqrt{m_{e}\left( r_{\Sigma }\right)
}\,t\,\right) ^{{2}/{3}}\longrightarrow 0
\end{equation}
because $ m_e(0)=0 $ and the wormhole disappears asymptotically,
which doesn't make sense physically, and this possibility is
discarded. Hence,
\begin{equation}
R_{\Sigma }\longrightarrow \left( r_{\infty
}^{3/2}+\frac{3}{2}\sqrt{m_{e}\left( r_{\infty }\right)
}\,t\,\right) ^{2/3}
\end{equation}
and the wormhole shell becomes comoving at late times.

We conclude this section with a comment on the energy conditions.
The strong energy condition for a $2$--dimensional perfect fluid
is $\sigma + P_{\Sigma}\geq0$ and $\sigma+2P_{\Sigma}\geq 0$: for
our wormhole, assuming a non--negative surface density $\sigma$
and hence $v< 0$, it is
\begin{equation}
\sigma +P_{\Sigma } =  \frac{\sigma }{2}+\frac{\gamma v_{t}}{4\pi
} \, \frac{1+v^2}{1-v^2}  > 0
\end{equation}
and
\begin{equation}
\sigma +2P_{\Sigma }=\frac{\gamma v_{t}}{4\pi } \, 
\frac{1+v^2}{1-v^2}  > 0 
\end{equation}
because $v<0$ and $v \rightarrow 0^-$ as $t \rightarrow \infty$, 
hence $v_t >0$.  The weak energy condition on the shell 
corresponds to $\sigma \geq
0 $ and $\sigma+P_{\Sigma} \geq 0$, while the null energy
condition is equivalent to $\sigma+P_{\Sigma} \geq 0$. Therefore,
the material on the shell satisfies the weak, strong, and null
energy conditions.

\section{The covariant conservation equation}

We can now solve the covariant conservation equation projected
along the $4$--velocity of the shell $u^{a}_{(\Sigma)}$
\cite{BarrabesIsrael91},
\begin{equation}
u_{\left( \Sigma \right) }^{a}\nabla _{b}\,S_{a}^{\,\,b}=-\left[
u_{\left( \Sigma \right) }^{\alpha }T_{\alpha }^{\,\,\beta
}n_{\beta }\right] \,.
\end{equation}
It is convenient to note that
\begin{eqnarray}
u^a_{(\Sigma)} \nabla _{b}\,S_{a}^{\,\,b} &= & 
 u_{\left( \Sigma \right) }^{a}\nabla _{b}\left[ \left(
\sigma +P_{\Sigma }\right) u_{a}^{\left( \Sigma \right) }u_{\left(
\Sigma \right) }^{b}\right] +u_{\left( \Sigma \right) }^{a}\nabla
_{a}P_{\Sigma } \nonumber\\
&&\nonumber\\
&= & -\nabla _{b}\left( \sigma \,u_{\left( \Sigma \right)
}^{b}\right) -P_{\Sigma }\nabla _{b}u_{\left( \Sigma \right) }^{b}
\end{eqnarray}
using the normalization $u_{\left( \Sigma \right)
}^{a}u_{a}^{\left( \Sigma \right) }=-1$ and its consequence
$u_{\left( \Sigma \right) }^{a}\nabla _{b}u_{a}^{\left( \Sigma
\right) }=0$. We now compute
\begin{equation}
\nabla _{b}\left( \sigma \,u_{\left( \Sigma \right) }^{b}\right)
=\frac{1}{\sqrt{\left| g_{\mid \Sigma }\right| }} \, \partial
_{b}\left( \sqrt{\left| g_{\mid \Sigma }\right| } \, \sigma
\,u_{\left( \Sigma \right) }^{b}\right) 
\,,
\end{equation}
where $g_{\mid \Sigma }= \gamma ^{-2} \, R_{\Sigma }^{4}\,\sin
^{2}\vartheta  $ is the determinant of the $3$--dimensional metric
$g_{ab\mid \Sigma }$, obtaining
\begin{equation}
\nabla _{b}\left( \sigma \,u_{\left( \Sigma \right) }^{b}\right)
=\frac{\gamma }{R_{\Sigma }^{2}} \, \partial _{t}\left( R_{\Sigma
}^{2}\sigma \right) \,=\gamma\,\frac{\dot{M}}{A_{\Sigma }}\,.
\end{equation}
Here $A_{\Sigma }\equiv 4\pi R_{\Sigma }^{2}$ is the area of the
shell and $M\equiv \sigma A_{\Sigma }$ is the mass of the material
located on the shell. Similarly, one obtains
\begin{equation}
\nabla _{b}u_{\left( \Sigma \right)
}^{b}=\gamma\,\frac{\dot{A}_{\Sigma}}{A_{\Sigma }}
\end{equation}
and
\begin{equation}
\left[ u_{\left( \Sigma \right) }^{\alpha }T_{\alpha }^{\,\,\beta
}n_{\beta }\right]= 2\,u_{\left( \Sigma \right) }^{\alpha
}T_{\alpha }^{\,\,\beta }n_{\beta } =-2 \gamma ^{2}\rho v \,.
\end{equation}
Putting everything together, we obtain the covariant conservation
equation in the form
\begin{equation} \label{conservation}
\dot{M}+P_{\Sigma }\dot{A}_{\Sigma }=-2\gamma \rho v A_{\Sigma
} \,.
\end{equation}
This formula is interpreted physically as follows: the quantity
$\rho\,v$ is the flux density of cosmic fluid onto the shell
caused by the relative motion of the shell with respect to the
background. The quantity $\rho vA_{\Sigma }$ is the flux of this
material;  the factor $2$ appears because there are two LTB
spacetimes joining at the shell. The Lorentz factor $\gamma$ is
due to the  Lorentz contraction caused by the radial motion of the
shell.

Eq.~\eqref{conservation} expresses the first law of thermodynamics
relating changes over a time interval $dt$
\begin{equation}
dM + dW =dQ_{\Sigma } \,,
\end{equation}
where $dM=\dot{M} dt $ is the variation of internal energy during
$dt$, $dW= P_{\Sigma }\, \dot{A}_{\Sigma } dt $ is a work term due
to the variation of the area of the shell, and $dQ_{\Sigma }$ is
the energy input due to the influx of cosmic fluid onto the shell.

\section{Discussion and conclusions}

We have obtained, and interpreted physically, an exact solution of
the Einstein field equations representing a wormhole shell joining
two identical LTB spacetimes which are dust--dominated. This
solution is similar to the wormhole solution of
Ref.~\cite{FaraoniIsrael05} obtained by generalizing the McVittie
metric, but there are important differences. First, we adopted the
no--accretion condition of Bondi \cite{Bondi} which forbids radial
flow of energy into the wormhole while 
Ref.~\cite{FaraoniIsrael05},
being interested in the effect of accretion, allows for radial
flow with the consequence that an imperfect fluid is needed in
order to obtain solutions in \cite{FaraoniIsrael05}. Here,
instead, we can consider a perfect fluid, the dust of classical
LTB solutions \cite{Lemaitre, Tolman, Bondi}. While in
\cite{FaraoniIsrael05} the conservation equation analogous to our
eq.~\eqref{conservation} has a right hand side consisting of two
terms, one due to the relative motion between shell and cosmic
substratum, and another due to accretion, only the first term
appears in our case in which there is no radial flux.

An important result of \cite{MaedaHaradaCarr09} is that, contrary
to static wormholes, the null energy condition needs not be
violated for their cosmological and dynamical wormholes to stay
open; here we propose a different cosmological wormhole solution
made with material which satisfies the weak, null, and strong
energy conditions on the shell. In other words, the ``stuff"
necessary to keep this wormhole throat open does not need to be
very exotic. This feature motivates further studies of dynamical
wormhole solutions of the Einstein equations.


%
%

%

\begin{acknowledgments}
I.B. thanks the Fonds Qu\'eb\'{e}cois de la Recherche sur la 
Nature et les Technologies (FQRNT) for financial 
support and Bishop's University for the hospitality. V.F. is 
supported by the Natural Sciences and Engineering Research 
Council of Canada (NSERC).
\end{acknowledgments}

\end{document}